\documentclass[letterpaper, 10pt, conference]{ieeeconf}      

\IEEEoverridecommandlockouts                              
\usepackage{graphicx}
  \usepackage{here}
  \usepackage{array}
  \usepackage{color,amsmath,here}
  \usepackage{amssymb}
  \usepackage{enumerate}
  \usepackage{ascmac}
  \usepackage{tascmac}
  \usepackage{fancybox}
  \usepackage{float}
  \usepackage{amsmath}
  \usepackage{mathrsfs}

\newtheorem{definition}{Definition}
\newtheorem{proposition}{Proposition}

\newtheorem{theorem}{Theorem}

\newtheorem{lemma}{Lemma}
\newtheorem{corollary}{Corollary}
\newtheorem{remark}{Remark}

\newtheorem{proofof}{Proof of }

\definecolor{mygreen}{rgb}{0, 0.7, 0}
\definecolor{myyellow}{rgb}{0.7, 0.7, 0}
\definecolor{mypurple}{rgb}{0.42, 0, 0.84}

\newcommand{\req}[1]{(\ref{#1})}

\title{\LARGE \bf
Weak Resilience of Networked Control Systems
}

\author{Tomonori Sadamoto$^{1,2,3}$, Henrik Sandberg$^{3,4}$, Bart Besselink$^{4}$, Takayuki Ishizaki$^{1,3}$\\Jun-ichi Imura$^{1,3}$, and Karl Henrik Johansson$^{3,4}$
\thanks{
$^{1}$Department of Mechanical and Environmental Informatics, Graduate School of Information Science and Engineering, Tokyo Institute of Technology; 2-12-1, Meguro, Tokyo, Japan:}%
\thanks{
\hspace{-3mm}{\tt\footnotesize \{sadamoto@cyb., ishizaki@, imura@\}mei.titech.ac.jp}}
\thanks{
$^{2}$Research Fellow of the Japan Society for the Promotion of Science}
\thanks{$^{3}$Japan Science and Technology Agency, CREST 4-1-8 Honcho,
Kawaguchi, Saitama, 332-0012, Japan}
\thanks{$^{4}$School of Electrical Engineering, Automatic Control, Royal Institute of Technology (KTH), SE-100 44 Stockholm, Sweden: }
\thanks{\hspace{-3mm}{\tt\footnotesize \{bart.besselink, hsan, kallej\}@ee.kth.se}}
}

\begin{document}

\maketitle
\thispagestyle{empty}
\pagestyle{empty}

\begin{abstract}
In this paper, we propose a method to establish a networked control
 system that maintains its stability in the presence of certain undesirable
 incidents on local controllers. We call such networked control systems \emph{weakly
 resilient}. We first derive a necessary and sufficient condition for
 the weak resilience of networked systems. Networked systems do not
 generally satisfy this condition. Therefore, we provide a method for
 designing a compensator which ensures the weak resilience of the compensated system. 
 Finally, we illustrate the efficiency of the proposed method by a power system example based
 on the IEEE 14-bus test system.
\end{abstract}

\section{Introduction}\label{sec:intro}
Many infrastructure and industrial processes, e.g., power networks \cite{giani2009viking, kundur1994power}, transportation networks \cite{bell1997transportation} and fabrication plants \cite{knapp2014industrial}, are integrations of computer-based cyber systems and physical processes. 
By emerging advanced technologies, the level of integration of the cyber and physical systems has intensified. Along with this, several challenging problems in control system design arise. 

Resilient system design is one of the most challenging problem for cyber-physical systems. 
The concept of resilient system design, which means control system design in an adversarial and uncertain cyber environment, has been introduced in \cite{rieger2009resilient}. Furthermore, in \cite{wei2010resilient}, the authors have discussed a conceptual property of resilient control systems. 
Moreover, 
in \cite{yuan2013resilient}, the authors have proposed resilient
controller design for cyber-physical networked systems under Denial of
Service (DoS) attacks which lead to severe time-delays and degradation of control performance. However, it is still an open problem to design resilient systems maintaining an acceptable level of operation or service in face of undesirable incidents on cyber systems, e.g., adversarial attacks and faults caused by human errors.

On the other hand, in \cite{sadamoto2014hierarchical}, the authors have
proposed a method for constructing systems whose stability is maintained
against any modification of local controllers, which stabilize local subsystems disconnected in the networked system.
In this method, we design a supervisory compensator such that the
compensated networked system has the property that the stability of the overall closed-loop system is guaranteed against any modification of locally stabilizing controllers.
However, no characterization of compensated networked systems having this property has been shown. 


This paper continues the research of \cite{sadamoto2014hierarchical} and establishes its connection to resilient control design, for the first time. 
First, we define weakly resilient networked systems such that the
overall closed-loop system maintains its stability in the presence of
any undesirable incidents on local controllers that maintain local
stability (to be defined in Section~\ref{sec:preliminary}). To clarify
the class of networked systems which are weakly resilient against
undesirable incidents on local controllers, we provide a necessary and
sufficient characterization of weakly resilient networked
systems. However, networked systems do not generally satisfy the shown necessary condition. Thus, we provide a design method to make a
given networked system weakly resilient. Finally, we show the efficiency
of the proposed system design through a power system
example based on the IEEE 14-bus test system \cite{zimmerman2011matpower}.

This paper is organized as follows. In Section \ref{sec:preliminary}, we introduce and characterize weakly resilient networked systems. In Section~\ref{sec:compensator}, we consider compensator design such that the networked system is weakly resilient.
In Section~\ref{sec:numsim}, we show the efficiency of the proposed system design through a numerical example. Finally, concluding remarks are provided in Section~\ref{sec:concl}.

{\bf Notation:}~ Denote the set of real numbers by $\mathbb{R}$, the set of complex
numbers by $\mathbb{C}$. Denote the $n$-dimensional identity matrix by
$I_{n}$, where we omit the subscript $n$ when no confusion occurs. 
For $\mathbb N := \{1,\ldots,N\}$, denote the block-diagonal
matrix having matrices $M_{1},\ldots,M_{N}$ on its diagonal by
${\rm dg}(M_{i})_{i\in \mathbb{N}}$. We omit the subscript $i\in
\mathbb{N}$ when no confusion occurs.
Given signals $x_1(t) \cdots x_N(t)$, denote $x(t) := [x_1^{\sf T}(t),
\ldots, x_N^{\sf T}(t)]^{\sf T}$, where we omit the time variable $t$ when no confusion occurs. Denote by $\Sigma:
u(t) \mapsto y(t)$ a finite-dimensional linear time-invariant system. 
Given $\kappa: y_1 \mapsto u_1$ and $\Sigma: \{u_1,u_2\} \mapsto \{y_1,y_2\}$, $(\Sigma, \kappa)$ denotes the (well-posed) interconnected system with the external input $u_2$ and external output $y_2$. For example, given 
\[
 \kappa: \left\{
\begin{array}{ccl}
\dot{\xi}&\hspace{-2.5mm}=&\hspace{-2.5mm}K\xi + Hy_1\\
u_1&\hspace{-2.5mm}=&\hspace{-2.5mm}M\xi
\end{array}
\right.
\]
and 
\[
\Sigma:
\left\{
\begin{array}{ccl}
\dot{x}&\hspace{-2.5mm}=&\hspace{-2.5mm}Ax + B_1u_1 + B_2u_2\\
y_1&\hspace{-2.5mm}=&\hspace{-2.5mm}C_1x\\
y_2&\hspace{-2.5mm}=&\hspace{-2.5mm}C_2x, \\
\end{array}
\right.
\]
$(\Sigma,\kappa)$ is the system described by
\[
(\Sigma,\kappa): \begin{array}{l}
\hspace{-1.6mm}
\left\{\hspace{-1mm}
\begin{array}{ccl}
\left[
\begin{array}{c}
 \dot{x} \\ \dot{\xi}
\end{array}
\right]
&\hspace{-2.5mm}=&\hspace{-2.5mm}
\left[\hspace{-1mm}
\begin{array}{cc}
 A & B_1M \\
 HC_1 & K
\end{array}\hspace{-1mm}
\right]\left[
\begin{array}{c}
 x \\ \xi
\end{array}
\right] + \left[\hspace{-1mm}
\begin{array}{c}
 B_2 \\ 0
\end{array}\hspace{-1mm}
\right]u_2\\
y_2&\hspace{-2.5mm}=&\hspace{-2.5mm}C_2x.
\end{array}
	\right.
\end{array}
\]
Denote the transfer matrix of the system $\Sigma: u\mapsto y$ by $\Sigma(s)$.
The $\mathcal L_2$-norm of a square integrable function
$v(t):\mathbb{R} \rightarrow \mathbb{R}^n$ is defined by 
$
\|v(t)\|_{\mathcal L_2} := \left(\int_{0}^{\infty} v^{\sf T}(t)v(t)
 dt\right)^{\frac{1}{2}}.
$
The $\mathcal{H}_{\infty}$-norm of a stable proper transfer matrix $G$ is defined by 
$
\|G(s)\displaystyle
 \|_{\mathcal{H}_{\infty}}:= \textstyle \sup_{\omega\in \mathbb{R}}\|G(j\omega)\|
$, where $\|\cdot\|$ denotes the induced $2$-norm.

\section{Weakly Resilient Networked Systems}\label{sec:preliminary}

\subsection{Definition of Weakly Resilient Networked Systems}\label{sec:definition}
\begin{figure}[t]
  \begin{center}
    \includegraphics[width=50mm]{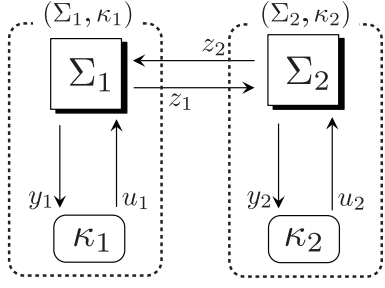}
    \caption{Overall closed-loop system $(\Sigma, \{\kappa_i\})$}
    \label{f01}
  \end{center}
\end{figure}

In this paper, for simplicity, we deal with a cyber-physical networked system composed of two subsystems, which represent dynamical processes in the physical world, and two local controllers, which represent cyber systems. We first give the dynamics of the cyber-physical networked system. For $i\in\{1,2\}$, the $i$-th subsystem dynamics are described by
\begin{equation}\label{subsys}
 \Sigma_i: 
  \left\{
   \begin{array}{rcl}
    \dot{x}_i &\hspace{-2mm}=&\hspace{-2mm} A_i x_i + J_{i}z_j + B_{i}u_i\\
    z_i &\hspace{-2mm}=&\hspace{-2mm} S_ix_i\\
    y_{i} &\hspace{-2mm}=&\hspace{-2mm} C_{i}x_i + D_{i}z_j,\\
   \end{array}
\right. \quad j\not=i,
\end{equation}
where $x_i \in \mathbb R^{n_i}$ is the state, $z_i \in \mathbb R^{p_i}$ is the subsystem interaction output, $u_i \in \mathbb R^{m_i}$ and $y_{i} \in \mathbb R^{q_i}$ are used for interconnection to the $i$-th local controller explained below. We suppose that $z_i$ and $y_i$ are measurable. The interconnection of $\Sigma_1$ and $\Sigma_2$ is given by
\begin{equation}\label{system}
 \Sigma: 
  \left\{
   \begin{array}{rcl}
    \dot{x} &\hspace{-2mm}=&\hspace{-2mm} A x + Bu\\
    y &\hspace{-2mm}=&\hspace{-2mm} Cx, \\
   \end{array}
\right.
\end{equation}
where
\begin{equation}\label{defABC}
 A\hspace{-0.5mm} = \hspace{-0.5mm}
 \left[\hspace{-1.5mm}
\begin{array}{cc}
 A_1& J_1 S_2\\
 J_2 S_1 & A_2
\end{array}\hspace{-1.5mm}
\right]\hspace{-0.5mm},
B \hspace{-0.5mm} = \hspace{-0.5mm}\left[\hspace{-1.5mm}
\begin{array}{cc}
 B_{1}& 0\\
 0 & B_{2}\\
\end{array}\hspace{-1.5mm}
\right]\hspace{-0.5mm}, C \hspace{-0.5mm} = \hspace{-0.5mm}
\left[\hspace{-1.5mm}
\begin{array}{cc}
 C_{1}& D_{1}S_2\\
 D_{2}S_1& C_{2}\\
\end{array}\hspace{-1.5mm}
\right]\hspace{-1mm}.
\end{equation}

For this networked system $\Sigma$, we consider designing local controllers to guarantee the stability of the whole closed-loop system. More specifically, for $i\in\{1,2\}$, we consider the $i$-th local controller generating $u_i$ from $y_i$ described by
\begin{equation}\label{lcon}
 \kappa_i: y_i \mapsto u_i, \quad i \in \{1,2\}.
\end{equation}
We denote the set of $\kappa_1$ and $\kappa_2$ by $\{\kappa_i\}_{i \in \{1,2\}}$. We omit the subscript $i\in\{1,2\}$ if no confusion occurs. The entire system $(\Sigma, \{\kappa_i\})$ is shown in Fig.~\ref{f01}. In this paper, we suppose that there exists $\{\kappa_i\}$ stabilizing $(\Sigma, \{\kappa_i\})$.

The local controllers in a cyber environment are sometimes vulnerable to adversaries and may be drastically 
modified. For example, in power grids, a local controller stabilizing
the frequency of power grids may be modified by attackers to cause power
outages. In addition, the controllers may be misconfigured by human
errors, which result in malfunctions of the system. In the face of such
undesirable incidents, the whole networked system is required to
maintain an acceptable level of operation, or at least preserve the
stability of the networked system.




As an example of undesirable incidents, let us suppose the following adversarial attack on the local controllers. Local controllers are modified by the attackers to achieve a desirable behavior of the system. However, since the modification with explicit consideration of the overall system dynamics is difficult, the attacker is supposed to focus on the dynamics of the local closed-loop system $(\Sigma_i, \kappa_i)$ as shown in Fig.~\ref{f01}, i.e., neglecting the interconnection to $\Sigma_j, j\neq i$. 
Furthermore, to avoid the detection of attacks as much as possible, we suppose that attackers do not destroy the stability of the local closed-loop system. 
In view of this, we consider attacks that preserve the stability of
local closed-loop systems $(\Sigma_i, \kappa_i)$.
In this setting, we define networked systems whose overall stability is
guaranteed against any adversarial attacks in the above class as follows:


\vspace{1mm}
\begin{definition}\label{def1}
For each $i\in\{1,2\}$, consider $\Sigma_i$ in \req{subsys} and
 $\kappa_i$ in \req{lcon}. Define $\Sigma$ in \req{system} and the set of locally stabilizing controllers as
 \begin{equation}\label{def_setK}
 {\cal K}_i := \{\kappa_i| (\Sigma_i, \kappa_i)~\mbox{is stable}\}
 \end{equation}
 for $i\in \{1,2\}$. The system $\Sigma$ is said to be {\it weakly resilient} if $(\Sigma, \{\kappa_i\})$ is
 stable for any $\kappa_i \in \mathcal K_i, ~ i\in\{1,2\}$. 
\end{definition}
\vspace{1mm}

The reason why we adopt the terminology {\it weak resilience} for this condition is that there does not exist a locally stabilizing
controller that destabilizes the overall system $\Sigma$. Hence, weak resilience, in this sense, appears to be a minimum requirement for
the resilience of networked systems.

In the next subsection, we will provide a characterization of weakly resilient networked systems.

\subsection{Characterization of Weakly Resilient Networked Systems}\label{sec:characterization}

In this subsection, we show a necessary and sufficient condition for the resilience of networked systems in the sense of Definition \ref{def1}.
For simplicity, we assume that the input and output signals of $\Sigma_i$ in \req{subsys} are scalar, i.e., 
\[
 z_i \in \mathbb R, \quad u_i \in \mathbb R, \quad y_{i} \in \mathbb R, \quad i\in\{1,2\}.
\]
In this setting, we give the following theorem:

\vspace{1mm}
\begin{theorem}\label{prop01}
 For each $i\in\{1,2\}$, consider $\Sigma_i$ in \req{subsys} and $\kappa_i$ in \req{lcon}. Define $\Sigma$ in \req{system}. Suppose $(A_i, B_{i})$ is controllable and $(A_i, C_{i})$ is observable for each $i\in\{1,2\}$.  The system $\Sigma$ is {\it weakly resilient} if, and only if, $\Sigma$ is a cascade system, i.e.,
 \begin{equation}\label{claim_thm}
  J_iS_j = 0, \quad D_{i}S_j = 0
 \end{equation}
 for either $i=1$ or $i=2$ and with $j\not=i$.
\end{theorem}
\vspace{1mm}
\begin{proof}
See Appendix.
\end{proof}
\vspace{1mm}

We emphasize that the cascade property of the system is not only a sufficient condition, but also necessary. In other words, if the system does not have any cascade realization, the system is not weakly resilient.
However, in general, networked systems are not necessarily cascade. Thus, in the next section, let us consider designing a compensator to make networked systems weakly resilient.


\section{Compensator Design for Weak Resilience}\label{sec:compensator}

\begin{figure}[t]
  \begin{center}
    \includegraphics[width=70mm]{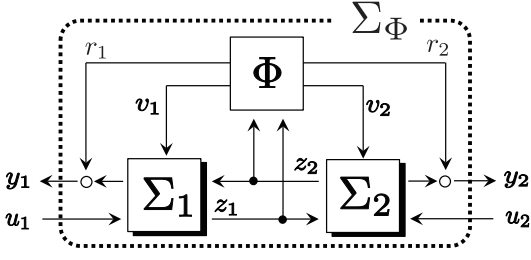}
    \caption{Compensated system $\Sigma_{\Phi} := (\Sigma, \Phi)$}
    \label{f03}
  \end{center}
\end{figure}

In this section, instead of $\Sigma$ in \req{system}, we deal with networked systems with additional input signals described as
\begin{equation}\label{system2}
 \Sigma: 
  \left\{
   \begin{array}{rcl}
    \dot{x} &\hspace{-2mm}=&\hspace{-2mm} A x + Bu + Rv\\
    y &\hspace{-2mm}=&\hspace{-2mm} Cx + r,\\
   \end{array}
\right.
\end{equation}
where $A$, $B$, and $C$ are defined as in \req{defABC}, and $v \in \mathbb R^{p}$ and $r\in \mathbb R^{q}$ are the additional input signals from the compensator introduced next. We suppose that $(A,R)$ is controllable. 


For this system, we consider designing a compensator described by
 \begin{equation}\label{compensator}
  \Phi: \left\{
\begin{array}{rcl}
 \dot{\phi}&\hspace{-2mm}=&\hspace{-2mm} \Lambda \phi + \Gamma z \\
 r &\hspace{-2mm}=&\hspace{-2mm} \Xi \phi\\
 v &\hspace{-2mm}=&\hspace{-2mm} \Theta \phi,
\end{array}
\right.
 \end{equation}
where $\phi\in\mathbb R^{\eta}$. 
Denote the compensated system by $\Sigma_{\Phi} := (\Sigma, \Phi)$. The network structure of this compensated system is shown in Fig.~\ref{f03}. 
In this setting, the following corollary follows from Theorem~\ref{prop01}:


\vspace{1mm}
\begin{corollary}\label{thm03}
Given $\Sigma$ in \req{system2}, consider $\Phi$ in \req{compensator}. Define the interconnected system $\Sigma_{\Phi} := (\Sigma, \Phi)$. Then, $\Sigma_{\Phi}$ is weakly resilient if, and only if, $\Sigma_{\Phi}(s)$ satisfies
 \begin{equation}\label{com_trans}
  \Sigma_{\Phi}(s) = 
   \left[
    \begin{array}{cc}
     C_{1}(sI - A_1)^{-1}B_{1}& \sigma(s) \\ 0 & C_{2}(sI - A_2)^{-1}B_{2}
    \end{array}
   \right]
 \end{equation}
 with a proper transfer function $\sigma(s)$, or $\Sigma_{\Phi}(s)$ has
 a similar lower-triangular form. 
\end{corollary}\vspace{1mm}


Next, we consider designing a compensator such that the transfer matrix $\Sigma_{\Phi}(s)$ has the form \req{com_trans}. As a related work, noninteracting control based on geometric control theory has been proposed in the literature, e.g., \cite{wonham1971status, willems1981disturbance}, where several off-diagonal elements of the transfer matrix are canceled. 
However, in general, the diagonal elements of the transfer matrix cannot be arbitrarily designed by the existing methods. Thus, existing methods do not enable us to construct $\Sigma_{\Phi}(s)$ having the form \req{com_trans} because the $i$-th diagonal element of
$\Sigma_{\Phi}(s)$ in \req{com_trans} must be $C_i(sI-A_i)^{-1}B_i$.


To overcome this difficulty,  in this paper, we consider designing a compensator by taking another approach, which was recently developed in \cite{sadamoto2014hierarchical}.
For simplicity, we assume that $D_{i} = 0$ in \req{subsys}. Note that
the system $\Sigma$ in \req{system2} is not a cascade. In this setting, we provide the following compensator on the basis of the
state-space expansion technique proposed in \cite{sadamoto2014hierarchical}: 


\vspace{1mm}
\begin{proposition}\label{one_compensator}
Given $\Sigma$ in \req{system2}, consider $\Phi$ in \req{compensator} with
\begin{equation}\label{dcom}
 \Lambda = \left[\begin{array}{cc}
 A_1& J_1S_2\\
      0&A_2
      \end{array}
\right] \hspace{-0.5mm}+ \hspace{-0.5mm}R\Theta,~ \Gamma =
 \left[\begin{array}{cc}
 0& 0\\
     J_2  & 0
       \end{array}
\right]\hspace{-1mm}, ~
\Xi =-{\rm dg}(C_i), 
\end{equation}
where $\Theta$ is given such that it stabilizes $A + R\Theta$. Then, $\Sigma_{\Phi} := (\Sigma, \Phi)$ is weakly resilient.
\end{proposition}
\vspace{1mm}

\begin{proof}
 The compensated system $\Sigma_{\Phi}$ is described by
\begin{equation}\label{clsd}
 \Sigma_{\Phi}: 
  \left\{\hspace{-1mm}
   \begin{array}{ccl}
  \left[
   \begin{array}{c}
    \dot{\phi} \\ \dot{x}
   \end{array}
  \right] &\hspace{-3mm}=&\hspace{-3mm}
  \left[
   \begin{array}{cc}
    \Lambda & \Gamma {\rm dg}(S_i)\\
    R\Theta & A \\
   \end{array}
  \right] \left[
   \begin{array}{c}
    \phi \\ x 
   \end{array}
  \right] + \left[
   \begin{array}{c}
    0 \\ B
   \end{array}
  \right]u\\
    y &\hspace{-3mm}=&\hspace{-3mm} -{\rm dg}(C_i)\phi + {\rm
     dg}(C_i)x. 
\end{array}
\right. 
\end{equation}
Taking the coordinate transformation $\chi = x - \phi$, we have
\begin{equation}\label{ex2}
  \left\{\hspace{-1mm}
   \begin{array}{ccl}
  \left[
   \begin{array}{c}
    \dot{\phi} \\ \dot{\chi}
   \end{array}
  \right] &\hspace{-3mm}=&\hspace{-3mm}
  \left[
   \begin{array}{cc}
    A+R\Theta & \Gamma {\rm dg}(S_i)\\
    0 & \mathcal A \\
   \end{array}
  \right] \left[
   \begin{array}{c}
    \phi \\ \chi
   \end{array}
  \right] + \left[
   \begin{array}{c}
    0 \\ B
   \end{array}
  \right]\hspace{-1mm}u\\
    y &\hspace{-3mm}=&\hspace{-3mm} {\rm dg}(C_i)\chi
\end{array}
\right.
\end{equation}
with
\[
 \mathcal A := \left[
\begin{array}{cc}
 A_1 & J_1S_2\\
 0 & A_2
\end{array}
\right].
\]
Hence, the transfer matrix $\Sigma_{\Phi}(s)$ has the form \req{com_trans}. Thus, $\Sigma_{\Phi}$ is weakly resilient.
\end{proof}
\vspace{1mm}

We note that the compensator $\Phi$ in \req{compensator} and \req{dcom}
relies on the use of two control inputs, i.e., $v$ and $r$. Existing
methods in noninteracting control generally only use the signal
$v$. However, the use of only $v$ does generally not allow for obtaining a
compensated system of the specific form \req{com_trans}. To this end,
the additional control input $r$ is exploited in the controller $\Phi$ in
\req{compensator} and \req{dcom}. 

Furthermore, in Proposition 1, it is shown that the whole closed-loop
system $(\Sigma_{\Phi}, \{\kappa_i\})$ preserves its internal stability
against any undesirable incidents on local controllers as long as $(\Sigma_i, \kappa_i)$ is stable.

Finally, we show a result on performance degradation of the whole
networked system under attacks on local controllers as follows. 
For simplicity, we take $\phi(0) = 0$ and the initial state of each
local controller as zero. We consider a closed-loop system $(\Sigma_{\Phi},
\{\kappa_i\})$ where $\Sigma_{\Phi}$ is given as \req{clsd}, and define
$x$ as the state of $\Sigma$ in this closed-loop system. Furthermore,
define $\chi$ as the state in the closed-loop system of \req{ex2} with local controllers of $\{\kappa_i\}$. 
In this setting, it follows that
\begin{equation}\label{perf}
 \|x(t)\|_{\mathcal L_2} \leq (1+\gamma)\|\chi(t)\|_{\mathcal L_2}
\end{equation}
for all $x(0) = \chi(0) = x_0 \in \mathbb R^n$ 
where $\gamma := \|(sI - (A+R\Theta))^{-1}\Gamma\|_{\mathcal H_{\infty}}$. 
Note that $\chi$ represents the state of the cascade system without interconnection from $\Sigma_1$ to $\Sigma_2$. Thus, we can see from \req{perf} that 
the performance of the overall closed-loop system is bounded by that of the local closed-loop systems. 
In general, it is not clear to what extent the performance of the whole closed-loop system is deteriorated under attacks on local controllers.
In contrast, the compensated system $\Sigma_{\Phi}$ has an advantage that the performance deterioration of the overall closed-loop system can be evaluated by that of the local closed-loop systems. 

\vspace{1mm}
\begin{remark}\label{remark_difference}
 In \cite{sadamoto2014hierarchical}, we have dealt with a similar
 compensator $\Phi$ but it made the transfer matrix $\Sigma_{\Phi}$ diagonal. In this case, it has been shown that the rank of $\Gamma$ in \req{compensator} coincides with the sum of the rank of $J_i$ for $i\in\{1,2\}$. Compared to this, $\Gamma$ in \req{dcom} is a lower-rank matrix.
Note that the low-rankness of $\Gamma$ has a direct relationship to the
 decay rate of Hankel singular values of $\Phi$. Thus, the compensator
 provided in this paper has a potential to be approximated by a
 lower-dimensional system as compared to the compensator considered in \cite{sadamoto2014hierarchical}. 
\end{remark}

\begin{remark}\vspace{1mm}\label{obs_based}
 Even if $z_i$ in \req{subsys} is not measureable, we can construct a compensator such that the compensated system is weakly resilient by using an observer as follows: We design an observer using 
\[
 w = Sx
\]
 as a measureable output signal of $\Sigma$ in \req{system2}. Define 
\begin{equation}\label{obs}
 O: \left\{\hspace{-1mm}
\begin{array}{ccl}
\dot{\hat{x}}&\hspace{-2.5mm}=&\hspace{-2.5mm}(A-HS)\hat{x}
+ {\rm dg}(B_i)u + Hw + Rv \\
 \hat{z}&\hspace{-2.5mm}=&\hspace{-2.5mm}\Gamma \hat{x}, \\
\end{array}\right.
\end{equation}
where $H$ is given such that $A - HS$ is Hurwitz. Let $\Phi$ be given by \req{dcom} using $\hat{z}$ instead of $z$. Then, $(\Sigma, \Phi, O)$ is weakly resilient.
\end{remark}

\section{Numerical Simulation}\label{sec:numsim}
\subsection{Power Network Model}\label{sec:powmodel}
\begin{figure}[t]
  \begin{center}
    \includegraphics[width=60mm]{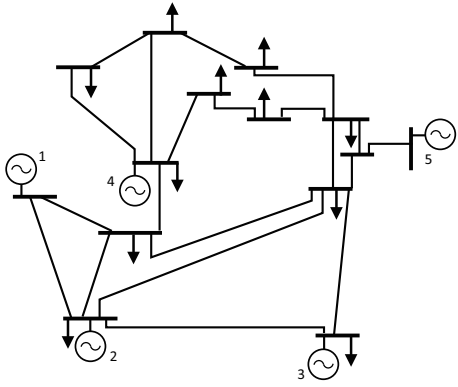}
    \caption{IEEE 14-bus test system}
    \label{pow14}
  \end{center}
\end{figure}

In this section, we show the efficiency of the proposed weakly resilient system design through a numerical example. We deal
with the IEEE 14-bus power test system provided by \cite{zimmerman2011matpower}, where the system involves five generators and 11 loads. 
The power system is shown in Fig.~\ref{pow14}. For $k \in \{1,\ldots, 5\}$, the $k$-th generator dynamics is described by
\begin{equation}\label{gener}
 G_{[k]}\hspace{-0.5mm}:\hspace{-0.5mm}\left\{
\begin{array}{l}
\dot{\zeta}_{[k]} = A_{[k]} \zeta_{[k]}
+b_{[k]} u_{[k]}
+b_{[k]} v_{[k]}
+b^{\tau}_{[k]} \tau_{[k]}\\
\delta_{[k]} \hspace{0.5mm} =  c \zeta_{[k]}, 
\end{array}\right.
\end{equation}
where the states of $\zeta_{[k]}\in \mathbb{R}^4$ represent the phase angle difference,
angular velocity difference, mechanical input difference, and valve position difference.
In addition, $u_{[k]} \in \mathbb{R}$ and $v_{[k]} \in \mathbb{R}$ are the angular velocity difference command, $\tau_{[k]}\in \mathbb{R}$ is the electric torque difference from the connected generators, and $\delta_{[k]} \in \mathbb{R}$ is the phase angle difference.
Furthermore, the system matrices in \req{gener} are given by
\begin{equation}\label{defAG}
\begin{array}{l}\hspace{-1.8mm}
A_{[k]}\hspace{-0.5mm}:=\hspace{-0.5mm}
\left[\hspace{-1mm}\begin{array}{cccc}
0 & 1 & 0 & 0 \\
0 & -D_{[k]}/M_{[k]} & -1/M_{[k]} & 0 \\
0 & 0 & -1/T_{[k]} & 1/T_{[k]} \\
0 & 1/K_{[k]} & 0 & -R_{[k]}/ K_{[k]}
\end{array}\hspace{-1mm}\right]\vspace{2mm}\\\hspace{-1.8mm}
b_{[k]}:= \frac{1}{K_{[k]}}e^{4}_{4},\quad 
b^{\tau}_{[k]} := \frac{1}{M_{[k]}} e^{4}_{2},\quad
c:= (e^{4}_{1})^{\sf T}, 
\end{array}
\end{equation}
where $e^n_i\in \mathbb{R}^{n}$ is the $i$-th column of $I_n$ and
$M_{[k]}$, $D_{[k]}$, $T_{[k]}$, $K_{[k]}$ and $R_{[k]}$ are an inertia constant, damping coefficient,
turbine time constant, governor time constant, and droop characteristic, respectively.
These parameters are randomly chosen from the intervals $[0.01,1]$, $[0.4, 11]$, $[0.01,0.02]$, $[0.03,0.7]$ and $[0.01,0.05]$, respectively. 
Note that the unit of all physical variables is [p.u.] unless otherwise stated. 
Furthermore, all loads are modeled as constant power loads, see \cite{zimmerman2011matpower}.

We give the interconnection structure among generators by
\begin{equation}\label{adm}
\tau=-Y\delta, ~ 
\end{equation}
where $\tau:= [\tau_{[1]}, \ldots, \tau_{[5]}]^{\sf T}$ and $\delta:= [\delta_{[1]}, \ldots, \delta_{[5]}]^{\sf T}$.
In \req{adm}, $Y$ compatible with the interconnection structure among generators is calculated by using MATPOWER \cite{zimmerman2011matpower}.

Finally, the first to third generators are clustered as the first
subsystem, and the others are clustered as the second subsystem.
Interconnencting these two subsystems, we have a system $\Sigma$ in
\req{system2} where the state variable is defined as $x =
[\zeta_{[1]}^{\sf T}, \ldots, \zeta_{[5]}^{\sf T}]^{\sf T}$, and input signals are defined as $u = [u_{[1]}, \ldots,u_{[5]}]^{\sf T}$ and $v = [v_{[1]}, \ldots, v_{[5]}]^{\sf T}$. 
Furthermore, the measurement signal is taken as the angle
differences, i.e., $y = [\delta_{[1]}, \ldots, \delta_{[5]}]^{\sf T}$. 
For the system matrices of $\Sigma$ in \req{system2}, $A$ is given by
\[
 A = {\rm dg}(A_{[k]}) - {\rm dg}(b_{[k]}^{\tau})Y(I_5 \otimes c) 
\]
where $\otimes$ denotes the Kronecker product. In addition, $B$, $R$ and $C$ are given as
the matrices compatible with $u$, $v$ and $y$.

\subsection{Demonstration of Compensator Design}\label{sec:demo}
\begin{figure}[t]
  \begin{center}
    \includegraphics[width=80mm]{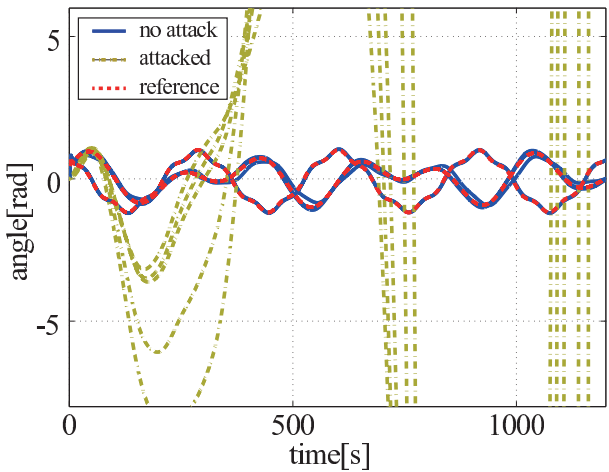}
    \caption{Transient responses of all generators in $(\Sigma,\{\kappa_i\})$ without using compensator $\Phi$}
    \label{fig:sim01}
  \end{center}
\end{figure}

\begin{figure}[t]
  \begin{center}
    \includegraphics[width=80mm]{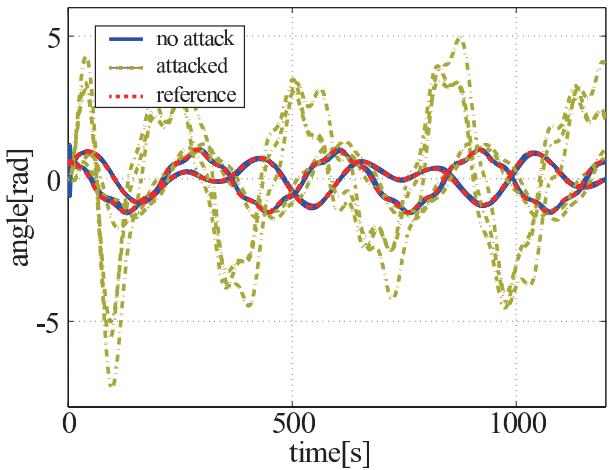}
    \caption{Transient responses of all generators in $(\Sigma_{\Phi},\{\kappa_i\})$}
    \label{fig:sim02}
  \end{center}
\end{figure}

\begin{figure}[t]
  \begin{center}
    \includegraphics[width=80mm]{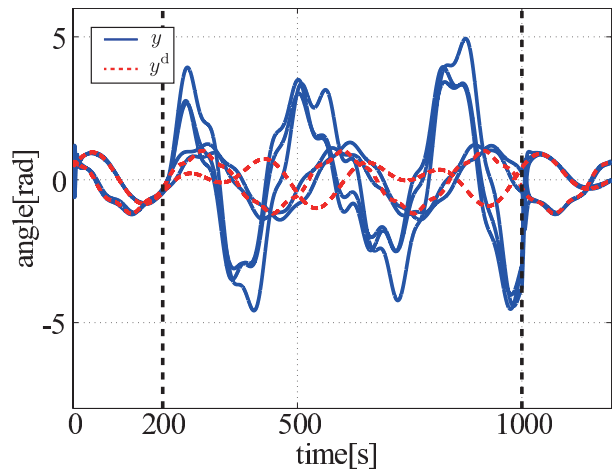}
    \caption{Demonstration of the power system operation under attacks on local controllers}
    \label{fig:sim03}
  \end{center}
\end{figure}

In this section, we show the efficiency of the compensator design for the power network given in the previous section.

First, we design the local controllers such that the power flow of the
whole closed-loop system is desirable when no adversarial attacks occur
in the local controllers. Since the power flow depends on the angle
differences among generators, we construct the local controllers such
that the angle difference $y\in\mathbb R^5$ tracks a given reference signal, denoted by $y^{\rm d} \in \mathbb R^5$. 
More specifically, given $\Sigma$ in \req{system2}, we consider an
augmented system whose states are $\dot{x}$ and the error between $y$ and
$y^{\rm d}$. For this augmented system, the local controllers
$\{\kappa_i\}$ in \req{lcon} are designed by LQR design techniques. 

To calculate the transient responses of the closed-loop system, we give
an initial state of the system and that of the controllers as zero.
Furthermore, we give the same reference signal in each subsystem, and each reference signal is taken as a random signal.

In Fig.~\ref{fig:sim01}, the blue solid (resp. red dotted) lines show the
transient responses (resp. reference signals) of the angle
differences of all generators when no attacks occur. 
We can see from this figure that the transient responses track the reference signals.
Furthermore, suppose that the local controllers are modified such that
the tracking performance of individual local closed-loop systems gets
worse, even though the local closed-loop systems are stable. 
In Fig.~\ref{fig:sim01}, the yellow dash-dotted lines depict the transient responses in this case. 
We can see from this figure that the instability of the closed-loop
system is induced by the attack on the local controllers. 

For the augmented networked system, we design $\Phi$ in \req{compensator} and \req{dcom} by minimizing $\gamma$ in \req{perf}, and construct a compensated system $(\Sigma, \Phi)$. 
The transient responses of the angle differences of all generators in
the case of $(\Sigma_{\Phi},\{\kappa_i\})$ are depicted in
Fig.~\ref{fig:sim02}, where the legends are the same as those in
Fig.~\ref{fig:sim01}. Furthermore, the (attacked) local controllers are the same as those shown above. 
From Fig.~\ref{fig:sim02}, even though the performance of
the closed-loop system becomes worse when the local controllers are
attacked, it should be emphasized that the stability of the whole system is preserved under attacks on local controllers by compensating the networked system by $\Phi$.



Finally, we numerically demonstrate the operation of the compensated power system under attacks on local controllers. To simulate this, we suppose a situation where local controllers are attacked while operating the whole system. 
We plot transient responses of the angle difference of all generators by the blue solid lines in Fig.~\ref{fig:sim03} during $t \in [0, 200)$.
Subsequently, we suppose that an attack occurs in the two local controllers at $t=200$ such that the tracking performance of individual local closed-loop systems gets worse. We can see from this figure around $t \in [200, 1000)$ that the stability of the whole system is preserved even though the tracking performance gets worse. Finally, we suppose that the controllers are recovered at $t=1000$. As a result, the tracking performance is recovered.
As shown in this numerical demonstration, the guarantee of the whole system stability against attacks on local controllers enables us to recover the controller while operating the whole power system.


\section{Conclusion}\label{sec:concl}
In this paper, we have proposed a method to establish a networked
control system that maintains its stability in the presence of certain
undesirable incidents on local controllers. We call such networked control systems \emph{weakly resilient}. To clarify the class of weakly resilient
networked systems,
we have provided a necessary and sufficient condition of weakly
resilient networked systems. However, networked systems do not
generally satisfy the necessary condition shown here in general. Thus,
we have provided a method for designing a compensator such that the compensated networked system
is weakly resilient. Finally, we have shown the efficiency of the proposed
method through a power system example of the IEEE 14-bus test system. 

In this paper, we have dealt with network systems composed of two subsystems, and shown a necessary and sufficient characterization of weakly resilient network systems. The generalization of this characterization to networked systems composed of an arbitrary number of subsystems is under investigation. 
Furthermore, we have shown a fundamental result of weakly resilient system design under undesirable incidents on local controllers preserving the stability of the local closed-loop system. 
The extension of this result to incidents destabilizing the system,
e.g., the stuxnet attack \cite{langner2011stuxnet}, is amongst the
topics of future works. 

\appendix

\begin{proofof}{\it Proof of Theorem 1:} 
The sufficiency is obvious. We show the necessity, i.e., $\Sigma$ in \req{system} is cascade if $(\Sigma, \{\kappa_i\})$ is stable for any $\kappa_i \in {\cal K}_i$, where $\Sigma_i$ and $\kappa_i$ are defined as in \req{subsys} and \req{lcon}, and ${\cal K}_i$ is defined as in \req{def_setK}. 

We first parametrize the $i$-th local closed-loop system $(\Sigma_i,
 \kappa_i)$ based on the Youla-parametrization in \cite{zhou1996robust}
 as follows. Let the input $u_i$ be composed of ${\bf u}_i$ and
 $\tilde{u}_i$ satisfying
\[
 u_i = {\bf u}_i + \tilde{u}_i, 
\]
where ${\bf u}_i$ is generated by a controller $\mbox{\boldmath$\kappa$}_i$ stabilizing $(\Sigma_i, \mbox{\boldmath$\kappa$}_i)$, i.e.,
\[
 \mbox{\boldmath$\kappa$}_i:
\left\{
\begin{array}{rcl}
 \dot{\xi}_i &\hspace{-2mm}=&\hspace{-2mm} (A_i+B_{i}F_i-H_iC_{i})\xi_i + H_iy_i\\
 {\bf u}_i &\hspace{-2mm}=&\hspace{-2mm} F_i\xi_i
\end{array}
\right.
\]
with $F_i$ and $H_i$ such that $A_i + B_{i}F_i$ and $A_i - H_iC_{i}$ are
 Hurwitz. In addition, $\tilde{u}_i$ is generated by a controller $\tilde{\kappa_i}: y_i \mapsto \tilde{u}_i$, which corresponds to the free parameter introduced below. The schematic depiction of the $i$-th local closed-loop system $\delta_i := (\Sigma_i, \kappa_i)$ is shown in Fig.~\ref{f04}. Let
 \[
 d_i := z_j,\quad j\not=i, 
 \]
and $\mathcal X_i = [x_i^{\sf T}, \xi_i^{\sf T}]^{\sf T}$, we have 
\begin{equation}\label{genplant}
 (\Sigma_i, \mbox{\boldmath$\kappa$}_i): \left\{
\begin{array}{rcl}
 \dot{\mathcal X}_i &\hspace{-2mm}=&\hspace{-2mm} {\bf A}_i \mathcal X_i + {\bf J}_i d_j + {\bf B}_i \tilde{u}_i\\
 z_i &\hspace{-2mm}=&\hspace{-2mm} {\bf S}_i \mathcal X_i \\
 y_i &\hspace{-2mm}=&\hspace{-2mm} {\bf C}_i \mathcal X_i  + D_i d_j, \\
\end{array}
\right.
\end{equation}
where 
\[
\hspace{-1.5mm}
\begin{array}{l}
 {\bf A}_i := 
\left[
\begin{array}{cc}
 A_i& B_{i}F_i\\
 H_i C_{i} & A_i + B_{i}F_i-H_i C_{i}
\end{array}
\right],\quad {\bf B}_{i} := 
\left[
\begin{array}{c}
B_{i}\\ 0
\end{array}
\right]\vspace{1mm} \\
{\bf J}_{i} := 
\left[
\begin{array}{c}
J_{i}\\ 0
\end{array}
\right]\hspace{-1mm},\quad
{\bf S}_{i} := \left[S_{i}~~ 0\right],\quad
{\bf C}_{i} := \left[C_{i}~~ 0\right]. 
\end{array}
\]
Hence, the closed-loop system $\delta_i$ can be parametrized as
\begin{equation}\label{def_delta}
  \delta_i(s; q_i) = {\bf \Sigma}_i^{dz}(s) + q_i(s){\bf
   \Sigma}_i^{\tilde{u}z}(s){\bf \Sigma}_i^{dy}(s), 
\end{equation}
where
\[
\begin{array}{l}
 {\bf \Sigma}_i^{dz}(s) = {\bf S}_i(sI - {\bf A}_i)^{-1}{\bf J}_i \vspace{1mm}\\
{\bf \Sigma}_i^{\tilde{u}z}(s) = {\bf S}_i(sI - {\bf A}_i)^{-1}{\bf B}_{i} \vspace{1mm}\\ 
{\bf \Sigma}_i^{dy}(s) = {\bf C}_{i}(sI - {\bf A}_i)^{-1}{\bf J}_i + D_i
\end{array}
\]
and $q_i(s) := \tilde{\kappa}_i(s)(1 - {\bf C}_i(sI-{\bf A}_i)^{-1}{\bf
 B}_i\tilde{\kappa}_i(s))^{-1}$. Note that there exists
 $\tilde{\kappa}_i(s)$ for any $q_i(s) \in {\mathcal RH}_{\infty}$. 
 Thus, we can regard $q_i(s) \in {\mathcal RH}_{\infty}$ as
 a free parameter. 
\begin{figure}[t]
  \begin{center}
    \includegraphics[width=50mm]{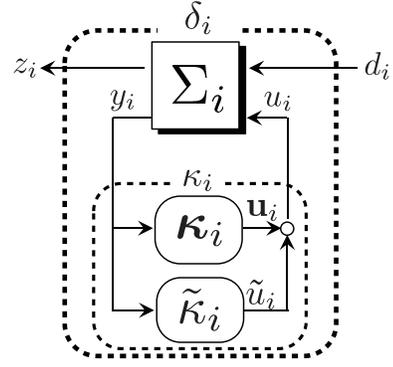}
  \caption{The $i$-th local closed-loop system}
  \label{f04}
  \end{center}
\end{figure}

\begin{figure}[t]
  \begin{center}
   \includegraphics[width=30mm]{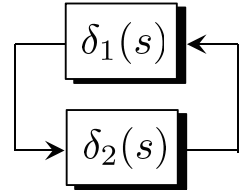}
  \caption{Networked system with two local controllers}
  \label{f05}
  \end{center}
\end{figure}

Next, we show a necessary condition on $q_i$ to guarantee the stability of the closed-loop system $(\delta_1, \delta_2)$ as shown in Fig.~\ref{f05} for any $q_i$ stabilizing $\delta_i(s;q_i)$.
Since $\delta_i$ is supposed to be stable by the assumptions of Theorem 1, it follows from the Nyquist
 stability theorem in \cite{zhou1996robust} that the interconnection $(\delta_1, \delta_2)$ as shown in Fig.~\ref{f05} is stable if, and only if, all roots of 
\begin{equation}\label{proof_roots}
 1 - \delta_1(s;q_1)\delta_2(s;q_2)=0
\end{equation}
 are in the open left half complex plane. Based on this fact, we show the following lemma: 

\vspace{1mm}
\begin{lemma}\label{lem01}
 The closed-loop system $(\delta_1(s;q_1), \delta_2(s;q_2))$ is stable for any $q_i$ stabilizing $\delta_i(s;q_i)$ only if
\begin{equation}\label{lem_claim01}
 f(q_1,q_2) := 1 - \delta_1(j\omega; q_1)\delta_2(j\omega; q_2)
\end{equation}
 is independent of $q_1(s)\in{\cal RH}_{\infty}$ and $q_2(s)\in{\cal RH}_{\infty}$.
\end{lemma}
\vspace{1mm}

\begin{proof}
 We first show that there exist $\omega \geq 0$ satisfying
  \begin{equation}\label{proof_lemlem}
   f(q_1, q_2) = 0,
  \end{equation}
 if there exist $\{q_i\}$ such that $(\delta_1(s;q_1), \delta_2(s;q_2))$
 is unstable, i.e., the roots of \req{proof_roots} are in the right half
 plane. Note that there exist $\{q_i\}$ such that the roots of
 \req{proof_roots} are in the open left half plane because there exist $\{\kappa_i\}$, which corresponds to $\{q_i\}$, stabilizing $(\Sigma, \{\kappa_i\})$. Hence, due to the continuity of roots, there exists $s = j\omega$ satisfying \req{proof_lemlem}. 
 
 Next, we show the claim by showing the contraposition. Namely, supposing that $f(q_1,q_2)$ depends on $q_i$, we show that there exist $\omega\geq 0$ and $q_i(s)\in {\cal RH}_{\infty}$ such that $(\delta_1,\delta_2)$ is unstable, i.e., $f(q_1, q_2) = 0$. 
 It follows from the fundamental theorem of algebra that there exist $\omega \geq 0$ and $\bar{q}_i \in \mathbb C$ satisfying
\[
 1-\bar{\delta}_1\bar{\delta}_2 = 0
\]
where $\bar{\delta}_i \in \mathbb C$ is defined as 
 \[
  \bar{\delta}_i :=  {\bf \Sigma}_i^{dz}(j\omega) + \bar{q}_i{\bf \Sigma}_i^{\tilde{u}z}(j\omega){\bf \Sigma}_i^{dy}(j\omega), \quad \bar{q}_i \in \mathbb C.
 \]
 Hence, it suffices to show that there exists a function $q(s) \in {\cal RH}_{\infty}$ satisfying $q(j\omega) = \bar{q}$ for a given $\bar{q} \in \mathbb C$ and $\omega \geq 0$.
 Give
\[
 q(s) = k\left(\frac{s - a}{s + a}\right)^2
\]
with $k > 0$ and $a > 0$. Since the $q(s)$ is an all-pass
 filter rotating $2\pi$[rad] while having gain $k$, there exist $k$ and
 $a$ satisfying $q(j\omega) = \bar{q}$. This completes the proof.
\end{proof}
\vspace{1mm}

It follows from Lemma~\ref{lem01} that we have
\begin{equation}\label{prflem}
 \frac{\partial f}{\partial q_2}(0,q_2) = 
{\bf \Sigma}_1^{dz}(j\omega){\bf \Sigma}_2^{\tilde{u}z}(j\omega){\bf
\Sigma}_2^{dz}(j\omega)=0, 
\end{equation}
where the first equality follows from the definition of $f$ in
 \req{lem_claim01} and \req{def_delta}. 
Thus, one of the transfer functions ${\bf \Sigma}_1^{dz}$, ${\bf \Sigma}_2^{\tilde{u}z}$, and ${\bf \Sigma}_2^{dz}$ is zero. 
Finally, we show \req{claim_thm} based on the analysis of three cases. 
\vspace{1mm}

\noindent
{\bf Case 1:} Suppose ${\bf \Sigma}_1^{dz}(j\omega) = 0$, i.e., 
\begin{equation}\label{prpr}
 {\bf S}_1(j\omega I - {\bf A}_1)^{-1}{\bf J}_1 = 0.
\end{equation}
Now, we introduce the following lemma:

\vspace{1mm}
\begin{lemma}\label{lem02}
Give $A \in \mathbb R^{n\times n}$, $B \in \mathbb R^n$ such that $(A, B)$ is controllable, and $C \in \mathbb R^{1\times n}$, $R \in \mathbb R^n$ and $D\in\mathbb R$. Then, 
 \begin{equation}\label{lem_claim02}
  C(j\omega I - (A+BF))^{-1}R + D= 0
 \end{equation}
holds for any $\omega \geq 0$ and $F \in \mathbb R^{1\times n}$ such that $A + BF$ is Hurwitz if, and only if, $D=0$, and $C = 0$ or $R = 0$.
\end{lemma}
\vspace{1mm}

\begin{proof}
 We only show the {\it only if} part. First, $D=0$ follows from \req{lem_claim02} because
 \[
 \lim_{\omega \rightarrow \infty} (j\omega I- A)^{-1} = 0.
 \]
 Next, we show $C = 0$ or $R = 0$. Note that $F = F - \tilde{F} + \tilde{F}$ for any $\tilde{F} \in \mathbb R^{1\times n}$. Since $(A,B)$ is controllable if and only if $(A + B(F - \tilde{F}), B)$ is controllable, no assumptions on stabilizability of $F \in \mathbb R^{1\times n}$ are required without loss of generality. It follows from the Laurent expansion that 
\[
 (j\omega I - A)^{-1} = \sum_{k=1}^{\infty}(j\omega)^{-k}A^{k-1}. 
\]
Hence, taking the derivative of \req{lem_claim02} with respect to $\omega$, we have
\begin{equation}\label{lemm}
 C(A + BF)^{k-1}R = 0 
\end{equation}
 for all $k \geq 1$ and $F\in \mathbb R^{1\times n}$. Taking $F = 0$, we have $CA^{k-1}R = 0$ for all $k \geq 1$. 
 Furthermore, let $k = 2$ in \req{lemm}. We have
\begin{equation}\label{lemm2}
 CBFR = 0 
\end{equation}
for all $F\in \mathbb R^{1\times n}$. Taking the derivative of \req{lemm2} with respect to $F$, we have $RCB = 0$. Let $k=3$ in \req{lemm}. It follows from $CBFAR = {\rm tr}(CBFAR) = {\rm tr}(RCBFA) = 0$ and $CA^2R = 0$ that
\[
 CABF + C(BF)^2R = 0
\]
Taking the derivative with respect to $F$ and $F = 0$, we have
\[
 RCAB = 0.
\]
Taking a similar procedure for $k > 3$, we have
\[
 RC[B, AB, \ldots, A^{n-1}B] = 0.
\]
Since the pair of $(A_i, B_i)$ is controllable, we have $RC = 0$, which
 is equivalent to $R=0$ or $C=0$. This completes the proof.
\end{proof}\vspace{1mm}

Note that ${\bf S}_1(j\omega I - {\bf A}_1)^{-1}{\bf J}_1$ in \req{prpr}
 is transformed into
\[
 [S_{1}~-S_{1}]\left(j\omega I \hspace{-1mm}-\hspace{-1mm}
 \left[\begin{array}{cc}
 A_1 + B_{1}F_1& -B_{1}F_1 \\
	0 &A_{1} - H_1C_{1}
       \end{array}
\right]\right)^{\hspace{-1mm}-1}\hspace{-1mm}
 \left[\hspace{-1mm}\begin{array}{c}
  J_{1}\\ J_{1}
       \end{array}
\hspace{-1mm}\right]\hspace{-1mm}.
\]
 Taking $F_1 = 0$ and using Lemma~\ref{lem02}, \req{prpr} shows that
 $S_{1} = 0$ or $J_{1} = 0$. Note that $S_1 = 0$ yields
 \req{claim_thm}. Thus, in what follows, we focus on the case that
 $J_{1} = 0$. Now, we have
 \[
  f(q_1,q_2) = 1 - q_1{\bf \Sigma}_{1}^{\tilde{u}z}D_1{\bf
 \Sigma}_{2}^{dz} - q_1q_2{\bf \Sigma}_{1}^{\tilde{u}z}D_1{\bf
 \Sigma}_{2}^{\tilde{u}z}{\bf \Sigma}_{2}^{dy}. 
 \]
 Hence, the independency of $f(q_1,q_2)$ from $q_1$ and $q_2$ is equivalent to 
\begin{equation}\label{prf_final0}
 {\bf \Sigma}_{1}^{\tilde{u}z}(j\omega)D_1{\bf \Sigma}_{2}^{dz}(j\omega) = 0
\end{equation}
and
\begin{equation}\label{prf_final1}
 {\bf \Sigma}_{1}^{\tilde{u}z}(j\omega)D_1{\bf \Sigma}_{2}^{\tilde{u}z}(j\omega){\bf \Sigma}_{2}^{dy}(j\omega) = 0. 
\end{equation}
We show \req{claim_thm} based on the case analysis as follows: 

{\bf Case 1a:} Suppose ${\bf \Sigma}_{1}^{\tilde{u}z}(j\omega) = 0$. Note that $B_1 \not=0$ follows because $(A_1, B_1)$ is controllable. Thus, ${\bf \Sigma}_{1}^{\tilde{u}z}(j\omega) = 0$ is equivalent to $S_1 = 0$, which yields \req{claim_thm}.

{\bf Case 1b:} Suppose $D_1 = 0$, which yields \req{claim_thm}.

{\bf Case 1c:} Suppose ${\bf \Sigma}_{1}^{\tilde{u}z}(j\omega) D_1 \not= 0$. Then, \req{prf_final0} yields that $S_2 = 0$ or $J_2 = 0$. 
Note that $S_2 = 0$ yields \req{claim_thm}. Thus, we focus on the case that $J_{2} = 0$. Then, ${\bf \Sigma}_{2}^{dy}(j\omega) = 0$ is equivalent to $D_2 = 0$, which yields \req{claim_thm}. 

\vspace{1mm}
\noindent
{\bf Case 2:} Suppose ${\bf \Sigma}_2^{\tilde{u}z}(j\omega) = 0$. Similar to case 1. 

\vspace{1mm}
\noindent
{\bf Case 3:} Suppose ${\bf \Sigma}_2^{dz}(j\omega) = 0$. Similar to case 1. 

\vspace{1mm}

\noindent
Therefore, we complete the proof of Theorem 1. 
 \hspace{\fill} $\blacksquare$
\end{proofof}
\bibliographystyle{IEEEtran}
\bibliography{reference.bib}
\end{document}